 \documentclass{aastex}

\newcommand{\myemail}{valerie@msfc.nasa.gov}

\slugcomment{accepted for publication in Astrophysical Journal}

\begin{document}

\title{BATSE Observations of Gamma-Ray Burst Tails}

\author{Valerie Connaughton\altaffilmark{1}}
\affil{NASA Marshall Space Flight Center AL 35812}
\altaffiltext{1}{work performed while National
Research Council Research Associate}
\altaffiltext{1}\myemail

\begin{abstract}
With the discovery of low-energy radiation appearing to come from the
site of gamma-ray bursts in the hours to weeks after the initial burst
of gamma rays, it would appear that astronomers have seen a 
cosmological imprint made by the burster on its surroundings.
I discuss in this paper the phenomenon of  post-burst
emission in BATSE gamma-ray bursts at energies traditionally
associated with prompt emission.  By summing the background-subtracted
signals from hundreds of bursts, I find that tails out to
hundreds of seconds after the trigger may be a common feature of long
events (duration greater than 2s), and perhaps
of the shorter bursts at a lower and shorter-lived level.  
The tail component appears
independent of both the duration (within the long GRB sample)
and brightness of the prompt burst emission,
and may be softer.
Some individual bursts have visible tails at gamma-ray energies and the spectrum
in at least a few cases is different from that of the prompt emission.
Afterglow at lower energies was detected for one of these bursts,  GRB~991216,
raising the possibility of afterglow observations over large energy ranges
using the next generation of GRB detectors in conjunction with sensitive
space or ground-based telescopes.
\end{abstract}

\section{Introduction}
Afterglow radiation at wavelengths from radio to X-rays
has now been seen in several gamma-ray bursts (GRB) 
detected with Beppo-SAX and the Burst and Transient Source
Experiment (BATSE) (e.g. ~\citep{nature97:costa,aa98:piro,nature97:jan}).  
This has allowed the unambiguous
determination of the cosmological distance scale of the bursters,
has led to the association of gamma-ray bursts with faint galaxies
rich in star-forming regions, and has opened up the possibility of using GRBs
as cosmological probes~\citep{5hgrb:lamb}.
It has also been a vindication of fireball models for the production of GRBs.
The fireball expands relativistically when
large amounts of bulk kinetic energy  are imparted to a 
small volume of plasma.  Gamma rays are released during the
fireball expansion and afterglow synchrotron radiation is anticipated
as a result of the interaction between this fireball and the material through
which it is moving.  

Afterglow radiation is detected when a previously unknown and fading source
is seen at a location consistent with the occurrence of a GRB.
It is clear that there is a frequency-dependent
temporal decay to the afterglow emission  and a variety of
afterglow temporal and spectral signatures have been seen.
The GRBs for which afterglows have been detected
encompass a wide range of brightnesses and
durations in prompt gamma-ray emission.
Indeed, there does not even appear to be a relationship between
the brightness of the initial GRB and the brightness or decay time of the
afterglow emission.
It is tantalizing to scientists that no afterglow has
been seen from the class of short GRBs defined as those lasting
less than 2 s.

Prompt emission at gamma-ray
energies was detected with BATSE~\citep{gro89:gjf}, and is
seen with the Beppo-SAX GRB Monitor
(GRBM)~\citep{asr91:frontera}, Ulysses~\citep{aas92:hurley},  HETE~\citep{aas99:vanderspek}
and other dedicated GRB
detectors.  Prompt X-ray 
emission can be seen by the Wide-Field Cameras (WFC) on SAX~\citep{asr95:jager}
 and HETE~\citep{aas99:vanderspek} or serendipitously
by the RXTE All-Sky Monitor~\citep{apj99:smith}.
The first and to-date 
most successful method used to pin down the afterglow
radiation from GRBs has been to slew the Narrow-Field Instruments (NFI) on the
Beppo-SAX satellite to observe a location from which the GRBM
and WFC  detected the prompt
emission.  The NFI
observations are typically made 6 to 8 hours after the GRB trigger.
A detection
of X-ray afterglow using the NFI provides a location on the sky accurate
enough for a search for optical afterglow using the most sensitive
telescopes.  These optical searches start about 1 day after the GRB trigger.
Optical searches have also been initiated as a result of follow-up 
observations by RXTE which typically start 3--4 hours after the GRB 
trigger~\cite{4hgrb:takeshima}.

Gaps exist between prompt and delayed emission observations and
their relationship is still unclear. Shock models for GRBs 
suggest that the initial
photons come from collisions of relativistic shells
of differing Lorentz factors initiated by some kind of central 
engine~\citep{apj94:rm},
while the afterglow is a result of the shock formed by
these shells colliding with an external medium~\citep{apj97:sp}.
It has been suggested in the case of GRB 970228 that 
if one extends the light curve of the X-ray afterglow back in
time, it joins smoothly with the prompt X-ray emission.  This
may indicate that some of the emission which appears to be part of
the prompt component
is consistent with being afterglow radiation~\citep{4hgrb:costa}.   
It is difficult to make
this argument for bursts in general:  the later peaks in bursts do not seem
to be significantly different from the initial ones,  and not all bursts
are multi-episodic.  The argument does open, however, the possibility
that some component in the late prompt emission might be related to
the afterglow.  One can then ask what does this component look like, 
when does it start and how long does it last?
~\citet{apj99:sp}
predict that the early afterglow of short GRBs will be separated from
the prompt emission by dozens of seconds, while in longer bursts the
afterglow will overlap the initial burst.  In their scenario, short bursts
are produced by internal shocks between thin shells, longer bursts
by thicker shells.
Recently,~\citet{apj00:fenimore} 
have proposed that the prompt
and afterglow gamma-ray emission may overlap if a shell of prompt photons is
decelerated by the external medium prior to being caught up by other
shells produced by the central engine.

In the work presented here, I explore the period of time from the BATSE GRB
trigger to 2000 s later 
using data between 20 keV and
2 MeV.   This time window is not usually probed in afterglow observations
because of the delays in obtaining good GRB locations and mobilizing
observing facilities.   
 The continuous BATSE data stream and the
development of a suitable background subtraction technique 
make possible the examination of long-term, low-level emission at
gamma-ray energies.
This technique will be described, followed by the results
of its application to 
a search for long-lived emission in  individual events and also in
 the combined signal from hundreds of events.

\section{BATSE measurements of GRB tails}
BATSE recorded continuously  with all 8 Large Area Detectors (LADs)
background data with 16 energy channel, 2.048 s time
resolution for the $\sim 80\%$ of the time
during which the detector voltages were activated
(they are disabled during South Atlantic Anomaly (SAA) passages), and the
spacecraft was in contact with a TDRSS satellite.  Over 
the $\sim 5600$~s spacecraft orbital
period, large, energy-dependent 
variations in background count rates are registered with the LADs
owing to passage through regions of differing 
magnetospheric activity.  Variations also occur in the cosmic ray 
component of the background rates because of changing rates of decay
of induced radioactivity in different magnetic fields.
A variation by a factor of about 1.4 in
count rates occurs in a typical orbit.
The triggering requirement for BATSE of $5.5 \sigma$ above the rates 
between 50 and 300 keV
registered during the previous 17~s in 2 or more of the 8 detectors 
in 64, 256, or 1024 ms ensures that the system responds to 
sudden increases in count
rates typical of the prompt gamma-ray emission of GRBs.  These increases 
(triggers) are
easily distinguished above variations in background rates.  In
contrast, GRB afterglows at
lower energies have typically been smooth on time-scales of thousands of seconds.
To ensure detection of a similar trend at higher energies,
careful background subtraction is required  to extract
any subtle signal over these time-scales in the BATSE system.  

\subsection{Background subtraction in the BATSE data}

In addition to the background variation 
pattern over each orbit, successive orbits are quite different 
owing to the Earth's rotation.  
The spacecraft orbit 
precesses, with a period of about 50 days, so that at the same time on two 
successive days, the spacecraft latitudes will have shifted.
The closest match in rates registered by the LADs occurs at times
 15 orbits before and 15 orbits after the orbit of interest.   
A background measurement is obtained by using the average of the count
rates from two orbits, each 15 orbital periods away from the
time of interest when a GRB triggers the instrument.

\begin{figure}
\plottwo{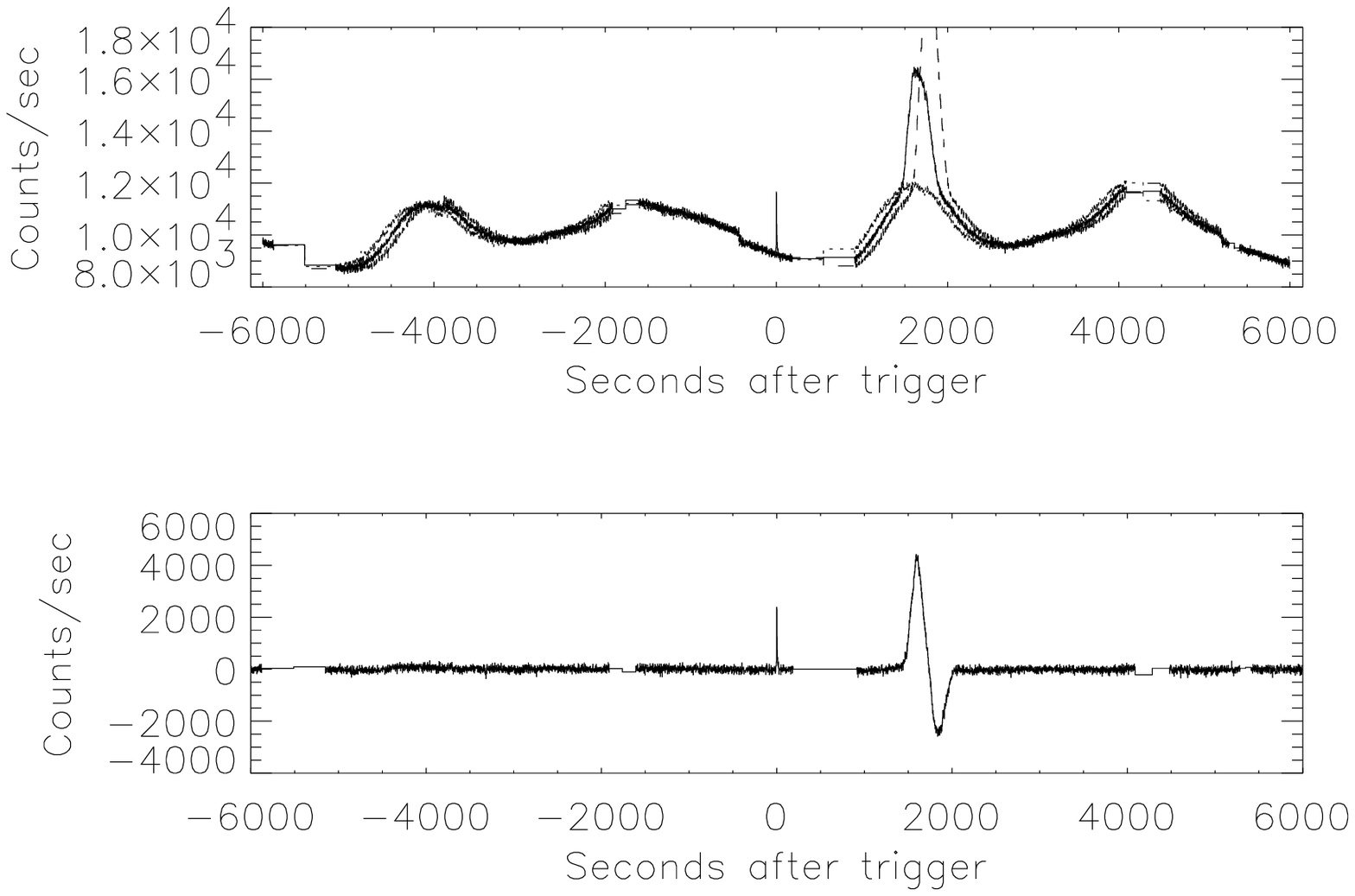}{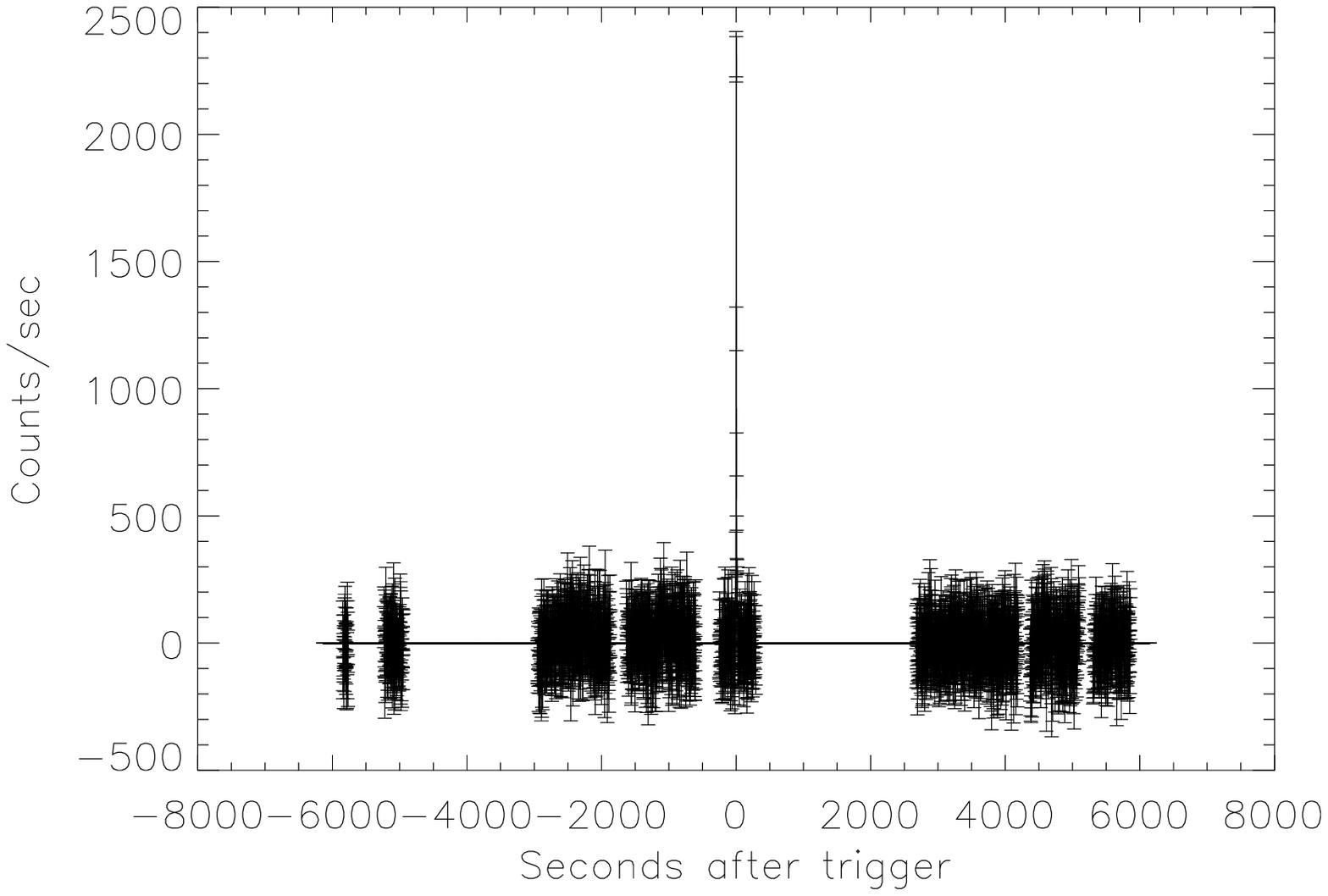}
\caption{3 superimposed BATSE LAD lightcurves showing the count rates 
from stretches of time covering two orbits on three successive days,
one centered on GRB trigger 6225, the others centered on times
15 orbits before and after the time of trigger which are to be used
to estimate the background before, during and after the burst.  The top
left plot shows the orbits separately, in the bottom left plot the average of 
the two background light curves is subtracted from the burst light curve.  
Telemetry gaps and regions of magnetospheric noise (between 1300 and 2000 s)
can be seen.  On the right hand side, the noisy regions have been removed 
and only times where the burst location is not Earth-occulted are included.  
The triggered event can be seen at $t=0$.  The errors shown
are statistical. \label{fig:6225}}
\end{figure}

Figure~\ref{fig:6225}~(left, top) shows the three superimposed 
lightcurves from a stretch of time covering two orbits on three successive
days.  One lightcurve is centered on GRB 970508, the
other 2 are centered on times 
15 orbits before and after the time of trigger of GRB 970508 and are to be used
to estimate the background before, during and after the burst.  
In Figure~\ref{fig:6225}~(left, bottom) the average of the background
curves has been subtracted.  It can be seen that in addition to the
triggered event at time $t=0$, there are noise events in
the orbits which are not constant from one to the next.  The most significant
of these discrepancies occur when the spacecraft exits or enters the SAA.  These
and other mismatches (electron events, phosphorescent spikes)
are flagged and removed in the analysis, resulting in the case of GRB 970508
in the profile shown in Figure~\ref{fig:6225}~(right).  The holes in the
light curve arise from telemetry gaps, disabling of the instrument during
passage through SAA, and the exclusion of data from 
times during which the burst position is occulted by the Earth.
In order to obtain a background measurement in this fashion
there must be no spacecraft re-orientation during the 3 days so that the
detector geometry relative to the burst position is constant. Telemetry
gaps must not occur at the time of the burst or at the corresponding
background orbit centers, and an additional requirement of no 
GRB triggers during the background orbits is imposed.  

The quality of this background fitting was assessed using test orbits
(which contain no triggered events)
centered at a time 5 orbital periods
after or before a GRB trigger.  
The test signal comprised the
sum of the count rates from the detectors which were triggered by the GRB
at times when the GRB position was visible to the spacecraft (i.e. not
Earth-occulted). 
Test background orbits were found on adjacent days separated in
time by 15 orbital periods, 
and the flatness of the resulting light curve after background
subtraction and cleaning was 
taken as a measure of the power of the technique.  It was found that the
light curves were flat, but that non-Poissonian errors  were present,
particularly in the lower energy channels 
and at high spacecraft geomagnetic latitudes.  
Slight
mismatches in the rates (probably due to spacecraft precession, and also
due to the noisier data at the peaks of the orbits) are more
severe at very high or very low latitudes than at the equator.  
A systematic error was added to the counting error, the 
magnitude of which depended on the latitude of the spacecraft.  
For each energy channel the systematic error  
was 10\% of the deviation of the background rate from the average 
(i.e. at the equator) background rate.  

\begin{figure}
\plotone{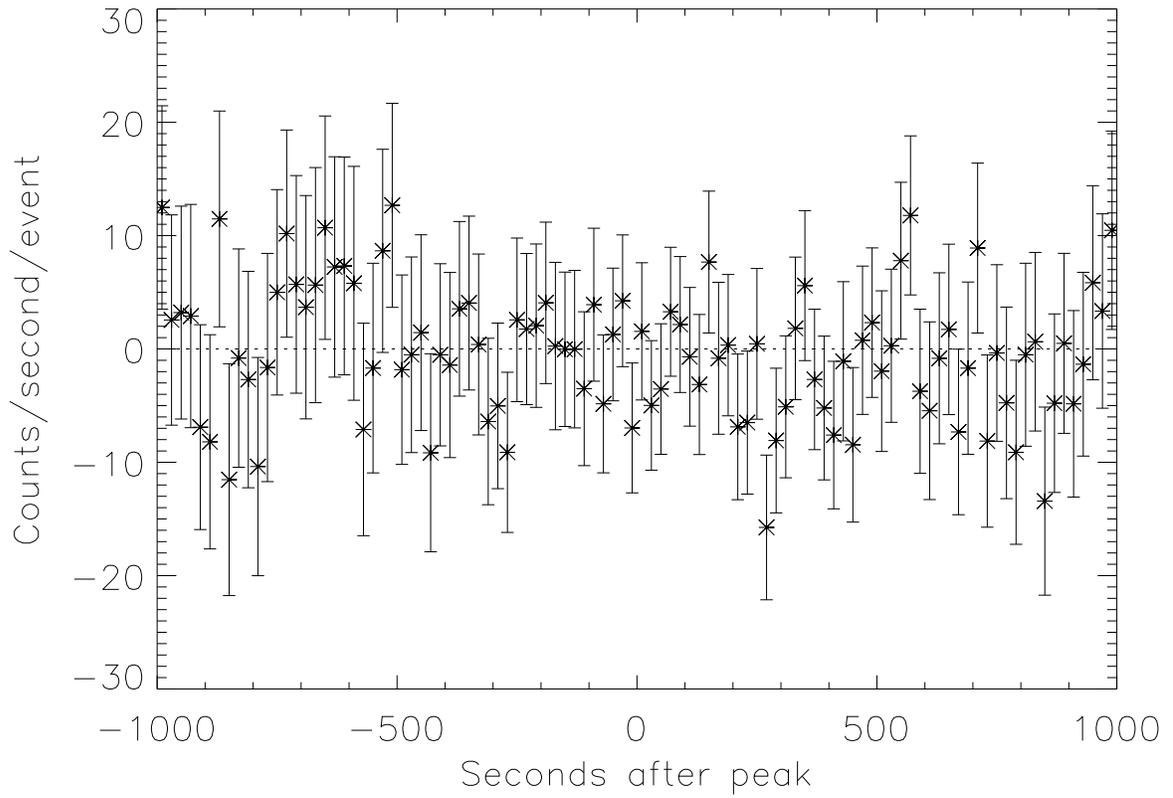}
\caption
{Lightcurve for 65 summed background-subtracted test
orbits  (summed energy channels).  The best
fit to the light curve is consistent with the zero-level dotted line.
\label{fig:bg_sys}}
\end{figure}

For an individual orbit, the sensitivity of this technique is estimated to be 
of the order of $10^{-9}$~erg~cm$^{-2}$s$^{-1}$, or an order of
magnitude lower than the BATSE trigger threshold.  The technique has
been successfully applied when examining
the possible presence of tails in the bright gamma-ray burst GRB~980923
~\citep{apj99:giblin}.  In the case of GRB~980923
the traditional method of background subtraction by linear interpolation
of intervals before and after the GRB failed owing to the duration of the
tail, the variations in background levels, and the faintness of the signal
in the tail.   This tail signal was more than two orders of magnitude fainter
than the bright peaks which precede it, and persisted for at least 200 s
whereas the main event stopped suddenly after 20 s.  Other bursts 
have been seen which exhibit bright spikes superimposed on
tail-like emission, but none with so dramatic a
transition as GRB~980923.  If this tail-like
emission is a feature of GRBs in general, then it occurs at a level which is
too faint to be visible by BATSE in most individual bursts, but might be
detectable in the combined signal from many events.  
In such an analysis, it is crucial that the addition of signal not be
hindered by an inadequate treatment of background count rates.
The robustness of the orbital background subtraction method developed here
to signal stacking was assessed by summing the background-subtracted
test orbits described above.  
Of 100 original test orbits, 35 were rejected for poor background 
fits - either visually or in a chi-squared test performed on data between
$-6000$ and $-4000$ s, and 4000 and 6000 s from the center of the test orbit.
Figure~\ref{fig:bg_sys} 
shows the summed light curve of the remaining 65 test orbits from
1000 s before to 1000 s after the aligned centers of the test
orbits.  The light curve is best fit by a 
a line of vertical offset $-0.5 \pm 0.7$~counts~s$^{-1}$~event$^{-1}$, consistent
with zero  (dotted line).

Having established the flatness of the residual light curve after background
subtraction and summation of light curves in the absence of GRB triggers,
background-subtracted signals for orbits centered on triggered GRBs
were summed, aligning the light curves at the peaks of the GRBs as 
~\citet{apj96:mitrofanov} did in their analysis of the asymmetry of burst peaks.  
The count rates in each 2.048~s time bin  in the orbits before and after
the GRB trigger during which the burst location was not Earth-occulted
and for which data exist for all three days 
were included in the summed light curve.
Ideally, the combined light curve would be a sum of all GRBs detected
by BATSE, but the event attrition rate was quite severe owing to the
data-intensive orbital background subtraction. 
BATSE triggered on 2365 cosmic GRBs between 1991 April and  1999 March.  
Of these, 526 occurred within 15 orbits of a spacecraft reorientation so
that at least one of the two orbits required for background subtraction
had a different detector geometry from that at the time of
the GRB trigger and could not be further analysed.
A further 595 were rejected because of the presence of a GRB trigger in
one of the two background orbits.
427 GRBs triggered in a telemetry gap (special burst data is
stored at these times but the continuous data stream
is unavailable), or a telemetry gap or SAA passage was
in progress at $t=0$ of at least one of the background
orbits.  These events were excluded owing to the lack of 
data around the time of the trigger.  
Background fits for 296 of the remaining 816 events were 
found to be unsatisfactory using a $\chi^2$
test so that in total,
520 BATSE GRBs were available for inclusion in the combined light curve.
     
Durations of GRBs detected with BATSE range from milliseconds to
hundreds of seconds.  The distribution of these durations is clearly
bimodal~\citep{apj93:ck} with a duration of 2s separating
the short from the long GRBs.
Efforts to associate the two classes with separate populations have
generally been unsuccessful,
although the bursts in the long part of the distribution do 
appear spectrally softer than the short ones~\citep{apj93:ck}.
The most recent distinction between the two classes has been the
lack of afterglow emission detected for the short ones.  This lack cannot
conclusively be claimed as an
absence of afterglow intrinsic to short GRBs, since no short bursts have been
seen by the WFCs on Beppo-SAX, and no sensitive
counterpart searches have, therefore, been
undertaken.    About 80\% of bursts seen with BATSE are in the long part
of the distribution, and this holds for the sample of 520 bursts analysed 
here.  The long and short bursts are treated separately in order to answer
the following questions:  Do long bursts in general show evidence for tail
emission such as that seen in GRB 980923 but at levels below what would be
 visible in individual events?  
Can any such tail emission be
seen at equally low levels in the combined signal from short GRBs?

\section{Results}
\subsection{Long Bursts}
Figure~\ref{fig:lin_ltails}
shows the combined light curve for the 400 long bursts.
The times are logarithmic and given relative to the aligned
peak, $t_0$, and the rates are summed above 20 keV and averaged by dividing
the total rates in a time bin by the number of bursts contributing
to that bin.  The time axis starts at $t_0 + 40$s. 
Bursts with precursor or successor  emission, defined
as outbursts which are separated from the time of the peak by longer than
the duration of the longest episode of the burst, are excluded.  This
exclusion is performed because these multi-episodic events can sometimes
have episodes of almost equal intensity so that deciding which peak to place
at $t=0$ becomes problematic.  

There are regions of the orbit following a trigger
which are not well-sampled - it is very likely, for example,
 that a burst
position will be occulted 2000 s after its detection.  Since only bursts
which trigger outside of gaps in the spacecraft telemetry are analysed,
the likelihood
of a telemetry gap or an SAA passage also increases with time.  The statistics
of the combined light curve are, therefore, better closer to $t=0$, partly
because the signal is larger, but
also because more bursts contribute to the averaged count rates.  
The ill-sampled light curve bins are eliminated by excluding
time bins where more
than 75\% of the burst positions from a sample of bursts
are occulted or occur in telemetry gaps.

\begin{figure}
\plottwo{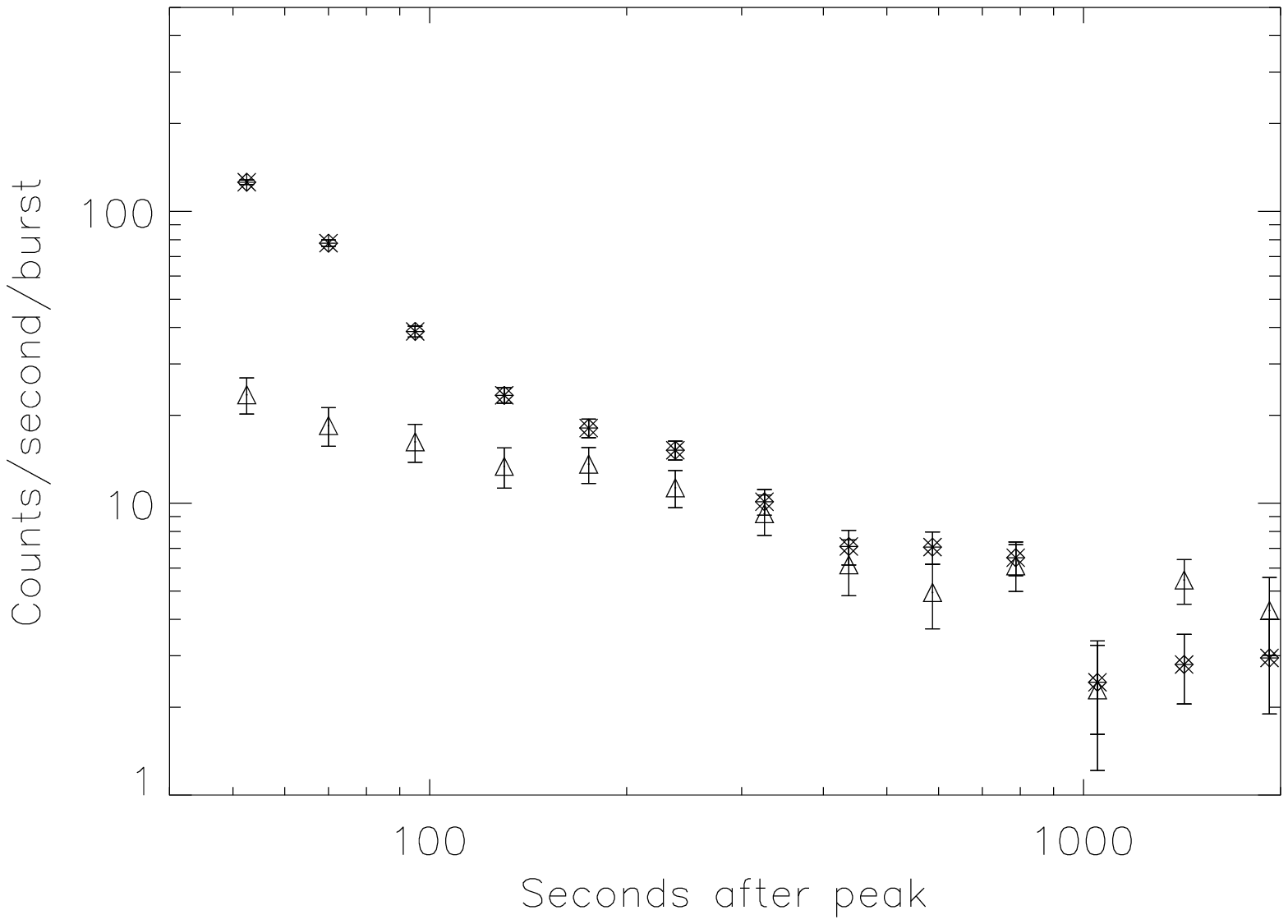}{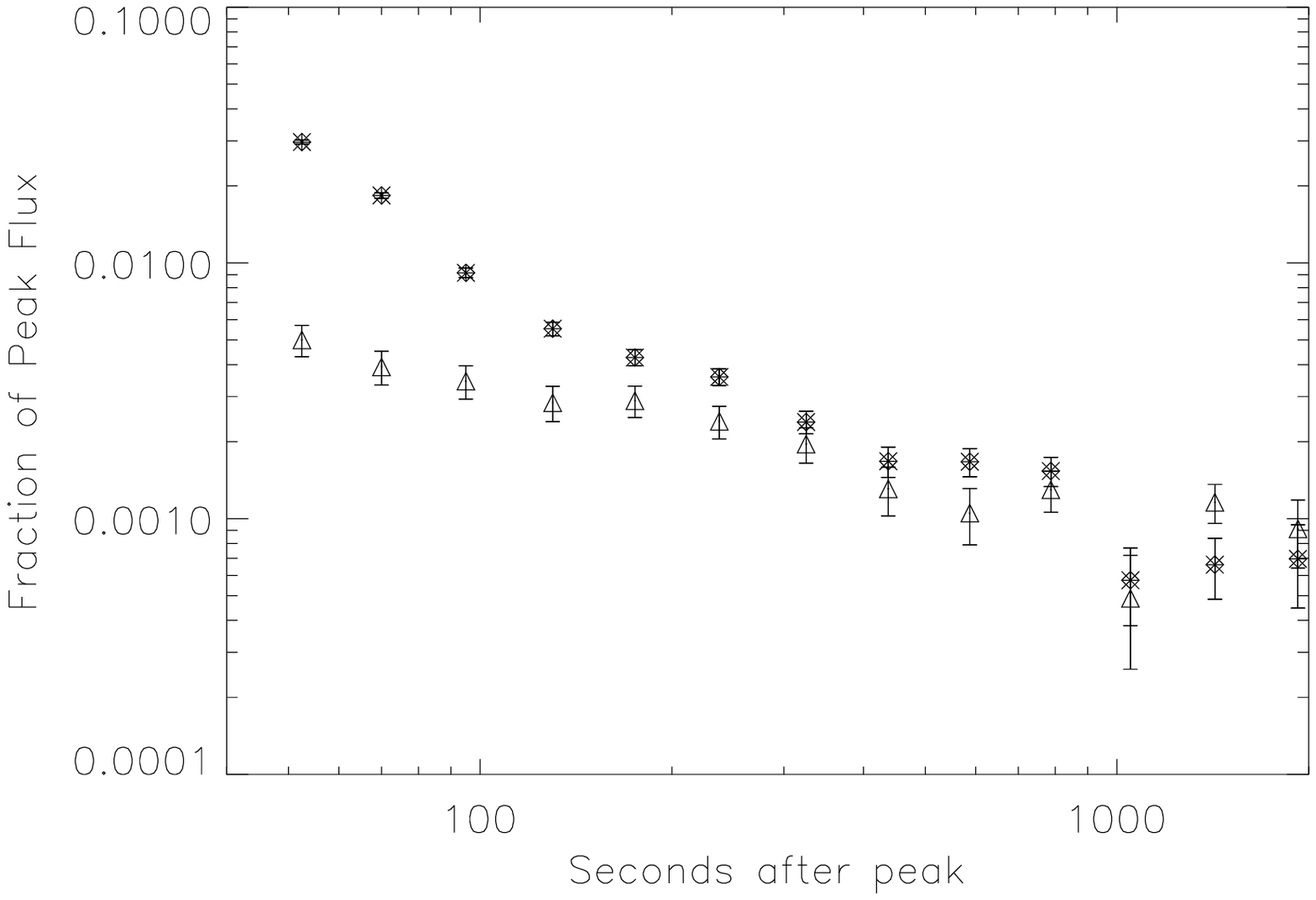}
\caption
{Lightcurve for summed background-subtracted long BATSE bursts
after peak alignment, with peak time suppressed.  Rates between 20
and 100 keV are shown for 400  bursts lasting longer than 2 s
(stars) and 200 medium bursts (triangles) with 30 s as the
dividing duration.  Left plot shows count rates per burst, right 
plot is the fraction of peak flux at a given time.~\label{fig:log_tails}}
\end{figure}

No tail-like excess is seen in the hundreds of seconds before the aligned
peak, and rates have returned to background following the ill-sampled
portions of the orbit which end at about t+4000s.  The period of interest
is that which occurs during and after the prompt emission.
It can be
seen that the summed light curve exhibits a tail-like low-level
of emission which extends beyond the duration of most GRBs.
The tail extends to at
least several hundred seconds after the aligned peak,
and may still be there in the
time bins between 1000 and 2000 s after $t_0$.
Although most long bursts last only tens of seconds, and those with
the longest measured durations are generally of the multi-episodic variety
which were excluded from this light curve,
it is reasonable 
to worry about the tail being a reflection of the duration
of the longest bursts in the sample.
The 400 long bursts were divided into two groups according to their
durations, and a medium duration subsample defined as 
shortest $50\%$ of long events was created which contains 200 bursts
with durations greater than 2 s but less than 30 s. 
Figure~\ref{fig:log_tails} starts at $t_0 + 40$ s  and shows on
a logarithmic scale the
count rates per burst in the long burst sample and the medium sample: 
the stars are the 400 bursts
longer than 2 s, the triangles are those 200 medium duration
bursts, whose main prompt emission is
deemed to be over by the time of the start of the plot.
Count rates per burst between 20 keV and
 100 keV are shown in the left plot, and are given
as a fraction of the peak count rates at $t_0$ on the right.
The tail is visible in both duration groups, although the 
count rates are higher for the longer
bursts until $t_0 + 200$ s since the longest bursts are still 
contributing prompt emission at this time.   
The contribution to the tail beyond this time is
independent of burst duration,  
suggesting a component is being detected which is distinct from the
prompt emission.  If the emission seen from the 200 medium duration bursts
is mostly tail emission from $t_0 + 40$s onward, then the average appearance 
of the tail is that shown by the triangles in Figure~\ref{fig:log_tails} and
can be fit by a power law of index $0.6 \pm 0.1$.  It is probable,
however, that like most burst and afterglow properties, this tail varies
considerably in individual bursts.

The spectrum of the emission in the combined light curve 
of Figure~\ref{fig:log_tails}  is characterized by
the ratio of counts detected in two adjacent energy channels.    
Figure~\ref{fig:hr} shows the evolution in time of the hardness of emission,
measured as the ratio of the counts between 50 and 100 keV to 
those between 20 and 50 keV sample,  starting at 10 s after $t_0$.
The stars show the summed complete long GRB sample exhibiting
a general softening
which steepens with time after about $t_0 + 200$ s until $t_0 + 600$.
The triangles indicate the trend for the summed medium
duration bursts where emission beyond $t_0 + 20$ s is flat at first and then
follows the softening trend seen in the longer bursts.  
The final point, which is the hardness ratios from the summed counts in
 the two bins at and above 1000 s in Figures~\ref{fig:lin_ltails} 
and~\ref{fig:log_tails}  indicates a hardening in
 both sets of bursts, though statistics are poor at these low flux levels.  This
hardening is seen in the subset of the 182 longest bursts (excluding the 
medium duration GRBs)  so it does not seem to be a contribution made solely
by medium duration bursts.  

Division of the 400 long bursts into three brightness-selected groups
indicates that the tail is measured in the summed light curves of  even
the dimmest group.   Figure~\ref{fig:bgroup} shows the intensity
of the tail for the 3 groups, the stars showing the brightest group (peak
flux $> 1.5$ photons~cm$^{-2}$s$^{-1}$ on a 1 s time-scale), the squares
pertaining to the dimmest bursts ($< 0.65$ photons~cm$^{-2}$s$^{-1}$ at
the peak intensity)
and the triangles to those of intermediate brightness.  Each group contains 
133 bursts.   
The intensity of the tail as a fraction of the peak
intensity (count rates at $t_0$)  for each subsample
is shown in the right hand plot of Figure~\ref{fig:bgroup}. 
Although the magnitude of the tail appears
largest in the bright group, the fraction of the peak
count rate in the tail is smaller than in the dimmer groups.  
The similarity of the tail in the three groups despite the variety of
bursts to which they pertain may lead to worry about the tail being
an artifact of the orbital background subtraction method.  It is clear from
the flatness of the summed test orbits in
Figure~\ref{fig:bg_sys}, however, that no such feature is
introduced by this methodology.

\subsection{Short Bursts}

\begin{figure}
\plotone{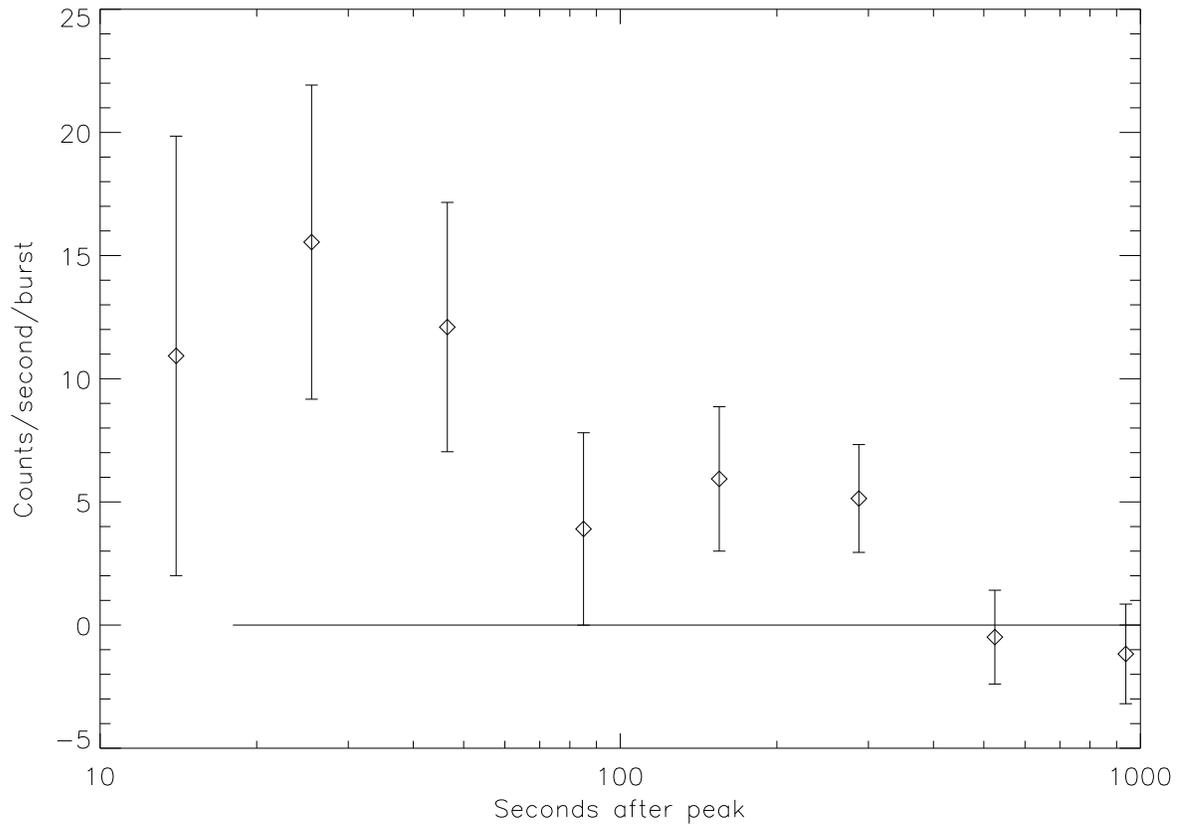}
\caption
{Lightcurve for 100 short ($<$ 1s) summed background-subtracted BATSE bursts
after peak alignment, with peak time suppressed.  \label{fig:lin_stails}}
\end{figure}

Figure~\ref{fig:lin_stails} shows the combined light curve for
the 100 short bursts - those with
durations less than 2 s - and starts at $t_0 + 10$s. 
Although the post-burst emission is not as prominent as in the summed
signals from the longer burst, an excess is apparent from $t_0 + 20$ s
out to $t_0 + 40$ s
and maybe even in the time period between $t_0 + 100$ s to $t_0 + 300$ s.
The level of tail emission per contributing burst is a factor of 3--4 smaller
than for the long bursts at this time, and the error bars are larger owing to
the smaller sample size, so that fitting this tail emission to a power-law or
other model of decay is not meaningful.
Likewise, the flux
is not large enough to compare its spectrum to the prompt emission or 
track any changes in hardness ratios throught the period of the excess.
This lightcurve is, however, less flat than
would be expected from the light curve obtained with the test orbits
shown in Figure~\ref{fig:bg_sys}. 
   
\subsection{Individual Bursts}
Long-lived gamma-ray emission is detected in
the combined signal from many GRBs in the long part of the bimodal
distribution and  does not seem to scale
with the brightness and duration of the prompt emission.
Some individual bursts have visible persistent low-level
emission following a much brighter epsiode,
and in the class of bursts known as FREDs (Fast Rise Exponential Decay)
the prompt emission decays gradually as a power-law in time.   
The orbital background subtraction method described above
reveals these types of long-lived emission with greater
 sensitivity and over longer time-scales than are first apparent.
The individual light curves from the bursts comprising the long-duration sample 
described above indicate that some have a more pronounced long-lived
component than others, thus contributing more to the combined tail,
without any obvious discriminant in the prompt
emission to indicate why this might be the case, 
and not all bursts appear to have a detectable long-lived component.
Likewise, afterglows at lower energies have been detected for 
a variety of GRB brightnesses and
durations,  but have not been seen for all GRBs for which
they were sought. Efforts to correlate the presence of afterglow radiation
with some property of the GRB prompt emission have not been successful.
The dramatic change in appearance and spectrum seen in GRB 980923 and reported
by ~\citet{apj99:giblin} suggests that
distinct emission components may be present in some GRBs.  If the 
long-lived, low-level emission seen in GRB 980923 is related to the afterglow
rather than the prompt emission of this burst, then one might expect
some correlation between the detection of gamma-ray tail emission and the
detection of lower-level afterglow emission.

Between 1996 July and 1999 December a total of
45 well-located GRBs were followed up in a search
for afterglow emission at X-ray, optical, infrared or radio wavelengths,
with varying degrees of depth of observation and delay times.  
Of this sample,  19 were detected by BATSE and
could be processed using the background-subtraction
methodology developed here.  
Although counting statistics and
background fluctuations in individual orbits limit the sensitivity to which
tails can be sought in individual bursts, each of these 19 events was
examined and categorized as being tailless, 
tailed, possibly tailed, or impossible
to categorize owing to poor or ambiguous background fits.  
Table~\ref{tab:ind} shows the list of
 obviously tailless events where the
prompt emission stops abruptly and the return to background is immediate,
and those where the presence of tailed, long-lived emission is apparent.  
The Table also indicates the presence of convincing afterglow
candidates detected at X-ray (X), optical (O), infrared (I) or radio (R) 
wavelengths.   A ``Y'' indicates a counterpart, a ``N'' implies observations
were made but no counterpart was found, and the field is left blank if
no counterpart searches were completed at a particular wavelength.
The afterglow observations
listed in Table~\ref{tab:ind} were taken from 
~\citet{ar00:jcr} who provide
a comprehensive list
of observations through 1999 July, and from circulars issued
to the Global Coordinates Network~\citep{gcn}.

The most dramatic gamma-ray tail is seen in GRB 991216.  Although the
tail emission is so faint compared to the peak emission 
that it cannot be seen without the background-subtraction method developed
here, it lasts at least 1100 s, at which time the position becomes
Earth-occulted, and may be faintly visible at 4000 seconds after the
occultation period ends.  The tail is so bright that an occultation
step can be seen in the background-subtracted data at the expected time.
The spectrum of the tail is significantly
different from the peak emission, and 
a more detailed analysis of this burst is
given in V. Connaughton et al. (in preparation).

Afterglows at all observed energies were seen for GRB 991216, as is also the
case for another two tailed events, GRB 980329 and GRB 980703.  
Only a weak limit can be set by the absence of any
counterpart to GRB 971024 because the follow-up
optical observations were not very deep, so that the trend of visible
tails existing for events with measured afterglows is not
seriously compromised by the lack of a counterpart for GRB 971024.
The non-detection of even an X-ray afterglow for GRB 970111 has always 
been puzzling given the brightness of the burst.  Any weak correlation
one might infer between gamma-ray tails and low-energy
afterglows is negated by the presence of a tail in the BATSE data
for GRB 970111.
The situation is further weakened  when one counts among the tailless events
the afterglow-rich GRB 971214, and especially GRB 970508 which
has such a convincing and much-observed afterglow light curve
measured over a long and broad-energy base line.  
 
No  correlation can, therefore, be implied between the presence of
gamma-ray tails and lower-energy afterglow detection using this 
small sample of GRBs.

\section{Discussion}

Emission in gamma rays from long GRBs seems to extend beyond 
what is first evident, at a level which is low compared to the prompt emission.
It appears to overlap the prompt emission and outlast it by more than 
several hundred seconds, following a smooth decay that can be
fit by a power law of index -0.6.  Such power-law decays have been
seen in the GRB afterglow detections at later times and lower
 energies, with indices varying between 1.1 and 1.9 (X-ray) and
1.1 and 2.4 (optical)~\citep{ar00:jcr}. 
The  results presented here were obtained by
summing the background-subtracted signals from 400 GRBs lasting longer
than 2 s
in order to establish the presence of gamma-ray tail emission in long bursts in
general.  This tail was also seen in
a medium duration sample comprising those 200 bursts
lasting between 2 and 30 s.
The similarity of the tail in the
summed light curve of the medium sample to that of the
long sample suggests a  
lack of correlation between the duration and magnitude of the tail emission and
the duration of the prompt GRB episode.  
The decay power law index of -0.6 
is an average value so that if the gamma-ray tail is related
to the afterglow at lower energies, one might expect
that the variations in decay indices 
seen at lower energies are also inherent in the
gamma-ray tails.   
In a study modeling the temporal decay of 40 GRBs,
 ~\citet{apj00:giblin} find a mean power-law of $2.05 \pm 0.51$,
which is only marginally
consistent with this work.  One might expect the
signal in the combined light curve to be dominated by flatter
decays at the later times explored in this analysis
(unless there is some dependence of the power-law decay index
on the initial brightness of the tail component) but the
average decay seen in this analysis is flatter than any single burst
modeled in ~\citet{apj00:giblin}.  This modeling of individual bursts
uses data which is closer to the peak emission than in the analysis presented
here, and it is possible that the tails in ~\citet{apj00:giblin} 
may represent a decaying of prompt
burst pulses rather than the shape of the gamma-ray afterglow
(except for the tail of 
GRB 980923 which has such a different spectrum from the main emission).

\begin{figure}
\plottwo{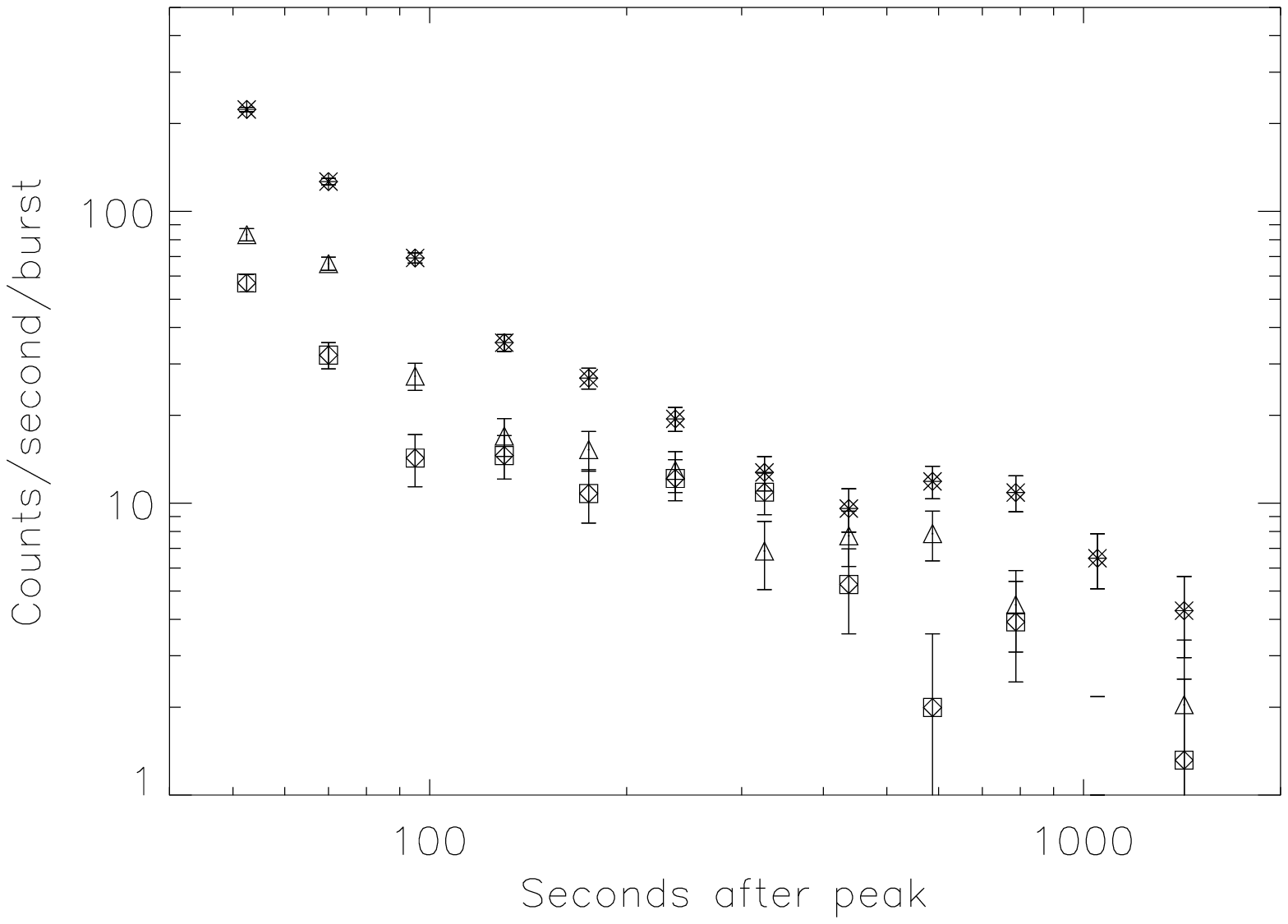}{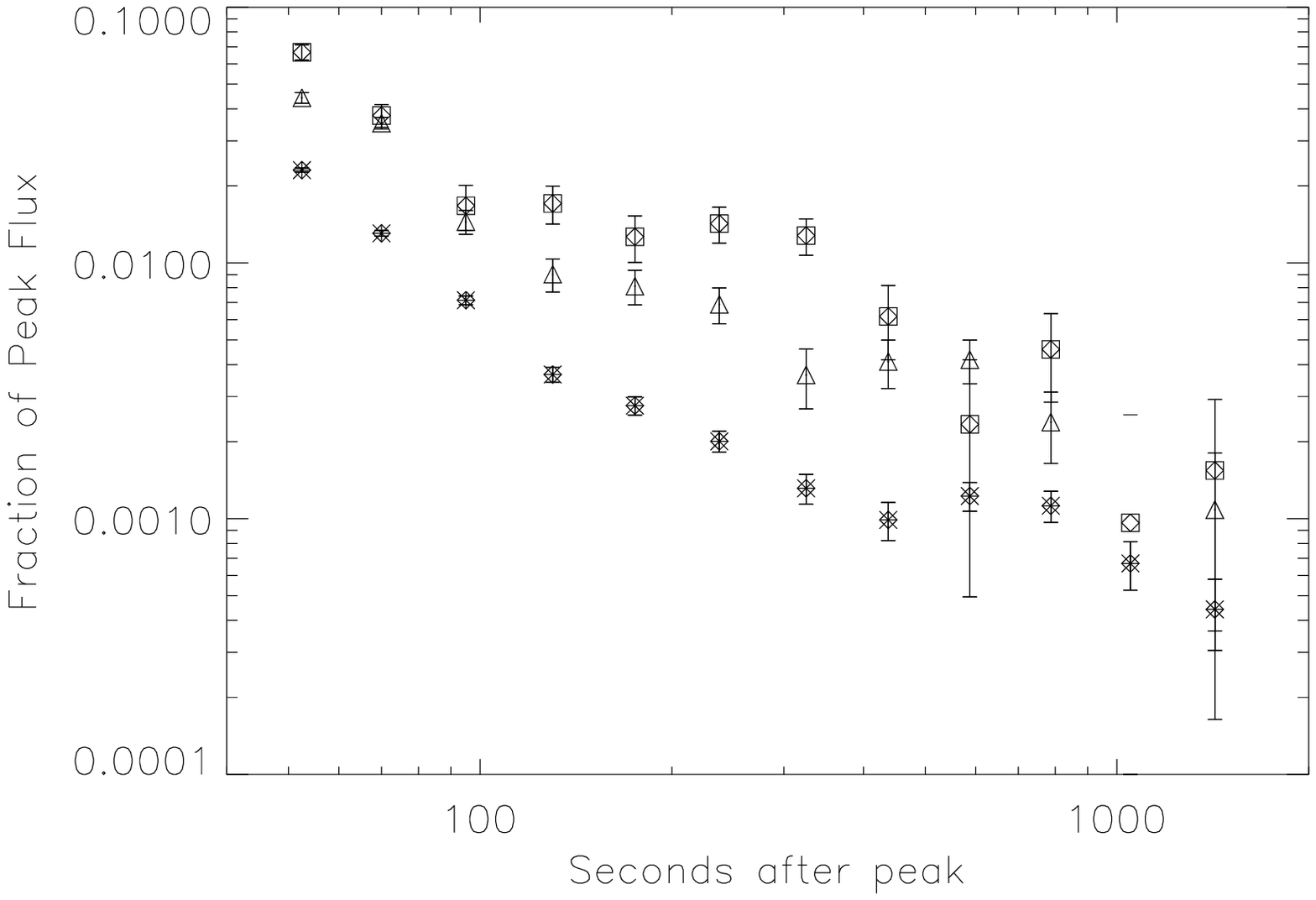}
\caption
{Three groups of long ($>$ 2s) bursts separated by the brightness
of their peak emission.  The left plot shows the count rate between 20 and 
100 keV per burst of the emission from $t_0 + 40$s onwards, the 
right shows the magnitude of this emission relative to the intensity at 
the peak.  The stars are the brightest bursts, the triangles intermediate, 
the squares the dimmest. \label{fig:bgroup}}
\end{figure}

By looking at tail emission as a fraction of
the peak flux, one can see that for the ensemble of medium 
duration bursts whose tails can be measured over a longer
time-line than the longer bursts, the emission at $t_0 + 40$ s
is about 4/1000  that at the peak, and 1000 s later it is an order
of magnitude lower yielding an approximate fluence  during
this time of 2 photons/cm$^{-2}$
above 20 keV, or $\sim 5 \times 10^{-7}$ erg/cm$^{-2}$ assuming a
 mean medium-duration GRB peak flux and a GRB-like power-law spectrum.
This compares with an estimate of the average medium duration burst
fluence of $\sim 4 \times 10^{-6}$ ergs/cm$^{-2}$ obtained by summing
the fluences between 20 and 300 keV for bursts in the sample using
values from the BATSE current catalog
({\it http://gammaray.msfc.nasa.gov}).
The 
observed brightness of a GRB, measured by its peak flux,
 is affected by its distance to the observer, 
but the wide range of luminosities implied by the measurements of just
a few GRB redshifts suggests a broad intrinsic luminosity function for 
the population of GRBs~\citep{5hgrb:lamb}.   By dividing the 400 long
bursts into 3 groups according to peak flux and examining their summed
light curves,  it was seen 
that although the  tail emission is most significant
in the brightest group,  
the difference is not as great as the
variation in peak fluxes between the ensembles.
The fact that there is some dependence of the
magnitude of the tail on the GRB peak brightness as seen in 
the left panel of Figure~\ref{fig:bgroup}
might be because
on average the dimmer bursts are further away. 
Figure~\ref{fig:bgroup} (right) shows, however, that 
as a fraction of peak count rates
the tail is brighter 
for the dimmer ensemble.
This implies a tail component which does not scale
directly with the intrinsic brightness of the prompt  GRB emission.
Whether this tail component is truly independent of the brightness
of the prompt emission cannot be inferred from this analysis,
but the intensity of the tail is clearly not directly
proportional to the intensity of the burst.
The average fluences of the prompt emission in the 3 brightness groups
obtained as above from the current BATSE catalog
are $1.5 \times 10^{-5}$, $3.2 \times 10^{-6}$, and $1.3 \times 10^{-6}$ 
ergs/cm$^{-2}$.  
The count-rate fluences of these groups, calculated between
$t_0 + 100$ and $t_0 + 1000$ s (to avoid including 
much prompt emission) calculated from Figure~\ref{fig:bgroup}
are 12000, 6000, and 5000 counts.  Assuming a similar spectrum in the
tails of the 3 brightness groups, the ratio of the fluence in the prompt
emission to the fluence in the tail does not follow a linear relationship.
The results appear consistent with the fluence in the tail scaling
with the square root of the fluence in the prompt emission 
to within the large errors associated with this crude estimate.

\begin{figure}
\plotone{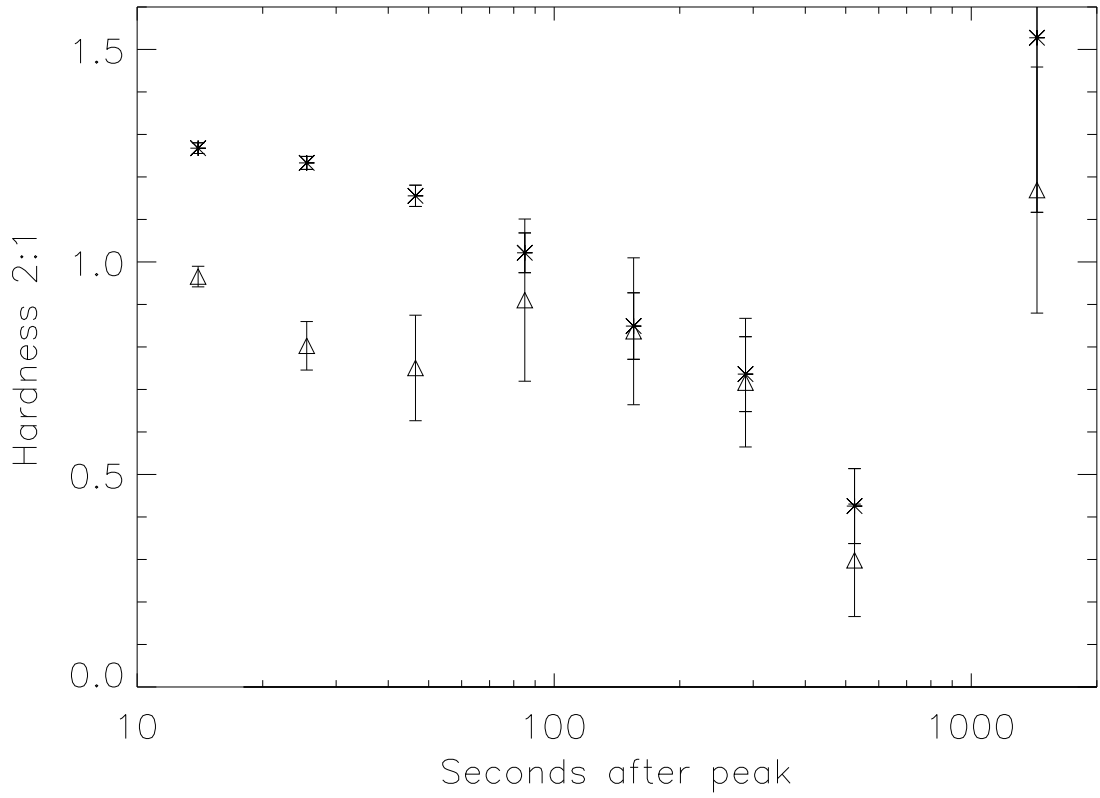}
\caption
{Hardness ratios (50-100 keV/20-50 keV) for long ($>$ 2s) bursts
(stars), bursts with durations between 2 and 30 s(triangles). 
\label{fig:hr}}
\end{figure}

\citet{apj99:giblin} 
and ~\citet{aa99:burenin} have found tail-like emission 
in the bright bursts GRB 980923 and GRB 920723, respectively.
A temporal decay in time with a power law index of $-0.69 \pm 0.17$
 was seen in the
case of GRB 920723, consistent with this analysis.  A much steeper index
of $-1.8 \pm 0.02$ 
 was measured for GRB 980923, illustrating the variety of decays
which may be contributing to the combined tail profile.
The fluence contained in the tails of bursts GRB 920723 and GRB 980923
is 20\% and 7\% of the total burst fluence, respectively, which agrees
with the estimated fluence in the tail between $t_0 + 40$ s and $t_0 + 1000$ s 
of 12\% of the average prompt GRB fluence for the medium duration sample.
One hundred seconds after the peak, the emission of GRB 920723 was 1/1000
that of its maximum flux,  that of GRB 980923 was a few thousandths of the peak
intensity.  
The intensity of this tail emission is considerably lower than that 
at the peak, consistent with what is seen in the analysis presented here,
but what really separates it from the main emission is a
sharp spectral transition.  The tail in each case is much softer than
the peak.
Clearly, the issue of spectral differences
between prompt and tail emission is important in establishing 
distinctions between the two components.  
Gamma-ray bursts in general
show a softening throughout their light curve~\citep{apj86:norris}
 so that any additional 
softening which is a signature of a separate component will be hard
to discern in the summed signal from many bursts.
With the overlap of prompt and tail emission of bursts of different durations
and spectral characteristics, it is difficult to separate 
these components.  There is some evidence for distinct spectral 
components in the main and tail emission which can be seen in
the evolution of the hardness ratio of the combined tail
signal shown in Figure~\ref{fig:hr}.  Softer emission is apparent in the
medium duration bursts until $t_0 + 100$ s compared to the entire set of
long bursts which contains events which are still in their prompt phase
during that time.  The flatness of the medium burst set from $t_0 + 20$ to
$t_0 + 300$ is consistent with what was seen in the tail of the individual
burst GRB 980923.  
Claims of distinct spectral identities for prompt and tail
emission are more convincing, however, 
in individual bursts where the spectral forms can be modeled rather than using
hardness ratios between two energy bands, and the transition between prompt
and tail emission is easier to discern.   

A sudden steepening and subsequent flattening of the spectrum seen at
$t_0 + 1000$ s in Figure~\ref{fig:hr} is more puzzling.   
This effect is seen in both the medium and 
the long burst data sets.  It is difficult to
quantify given the low fluxes involved, but qualitatively, at least,
this hardening may be consistent with the onset of an Inverse Compton 
(IC) component
which is anticipated by ~\citet{apj00:dermer} (in external shock GRB models);
by ~\citet{apj98:totani} (from synchrotron emission
by protons in the GRB jet to explain delayed high-energy
GRB emission such as that seen in GRB 940217~\citep{nature94:hurley}); and 
more recently by ~\citet{apj00:kumar} and ~\citet{apj01:sari}.  
Sari and Esin state  that this inverse Compton 
component might most easily be seen in the early afterglow at gamma-ray
and X-ray energies.   
Fast-cooling fireball scenarios
explored by ~\citet{apj01:sari} show a significant IC presence in the
gamma-ray spectrum 43 minutes after the trigger time.
While they expect the onset and importance of the
inverse Compton emission to vary from one burst to another, the hardness of
the summed rates from all the long ($> 2$s) bursts
between 1000 and 2000 s after their aligned peak might
be indicative of the onset of the IC frequency entering
the observable energy range of BATSE and the domination of
this IC component over the fading synchrotron tail.  
At later times in the afterglow, IC
emission has been reported by ~\citet{apj01:harrison} as 
contributing to the X-ray afterglow spectrum of GRB 000926 2 and
10 days after the GRB but no other afterglows have so far been
observed with this bump in their afterglow spectrum.   
   
In the analysis of the combined light curve,
the most convincing evidence for the distinct nature of the tail relative
to the prompt emission is the lack of
scaling of its duration and intensity with those of the early emission.
If the emission components are distinct, then their overlap in time is 
consistent theoretically with 
the deceleration model of ~\citet{apj00:fenimore}.  
This model suggests that bursts from fireballs with 
high Lorentz factors will exhibit two contemporaneous
emission components, one a result of an initial fireball-external
medium deceleration shock, the other the arising from
the internal shocks within the fireball.
If the Lorentz factor is low ($< ~ 1000$)
internal shocks occur prior to the first shell being decelerated
by the external medium, and only main prompt emission and
the post-burst afterglow will be seen.  
The deceleration depends, therefore, on the
initial Lorentz factor of the fireball, not on any
parameter such as the width of the shells (determining
duration of GRB peaks) or the total energy in the 
fireball (determining GRB intensity and fluence).
One signature of such a deceleration phenomenon is an overlapping tail
component which is different in nature from the main emission.  
~\citet{apj00:fenimore} postulate that this component will be smoother,
weaker and spectrally distinct from the main emission.
The analysis presented here finds a tail component which could 
result from this deceleration and  the possibility
that all long bursts exhibit this property  is not excluded.
A connection between GRB tail emission and the deceleration model has
a major impact on the energetics of GRBs.
The appeal of deceleration by an initial external shock is that it makes the
internal shocks more efficient by introducing a large difference in Lorentz
factors between leading and subsequent fireball shells.
Not so much of the bulk energy
remains to be dissipated following the prompt GRB episode, lowering the
total energy budget at the expense of a high initial Lorentz factor.

If in establishing the distinctness of the prompt and tail 
gamma-ray components detected here, one looks to lower energy 
afterglow observations rather than theory, the picture is also encouraging.
Afterglows at lower energies have been measured for both bright and dim
BATSE events, and for both medium and long-duration bursts, so that the
independence of a gamma-ray tail component of both duration and
brightness of the prompt emission is not unexpected if this
tail component is  related to the afterglow emission.

Not all bursts seem to have lower-energy
afterglow emission, and not all bursts have obvious gamma-ray tails.
The lack of correlation found here between the detection of afterglow
emission at X-ray, optical and radio energies and the presence
of a gamma-ray tail in individual events, 
however, suggests the relationship between the different
afterglow components may not be simple.  
It is not possible to draw conclusions based on
the small number of tailed and obviously tailless GRBs. 
The unambiguous
determination of the absence of a gamma-ray tail is often difficult given
the sensitivity of the BATSE detectors and the background-subtraction
technique.  Moreover, the searches for afterglows at lower energies are not    
always deep or timely enough to place severe limits on afterglow 
counterparts.   

\begin{figure}
\plotone{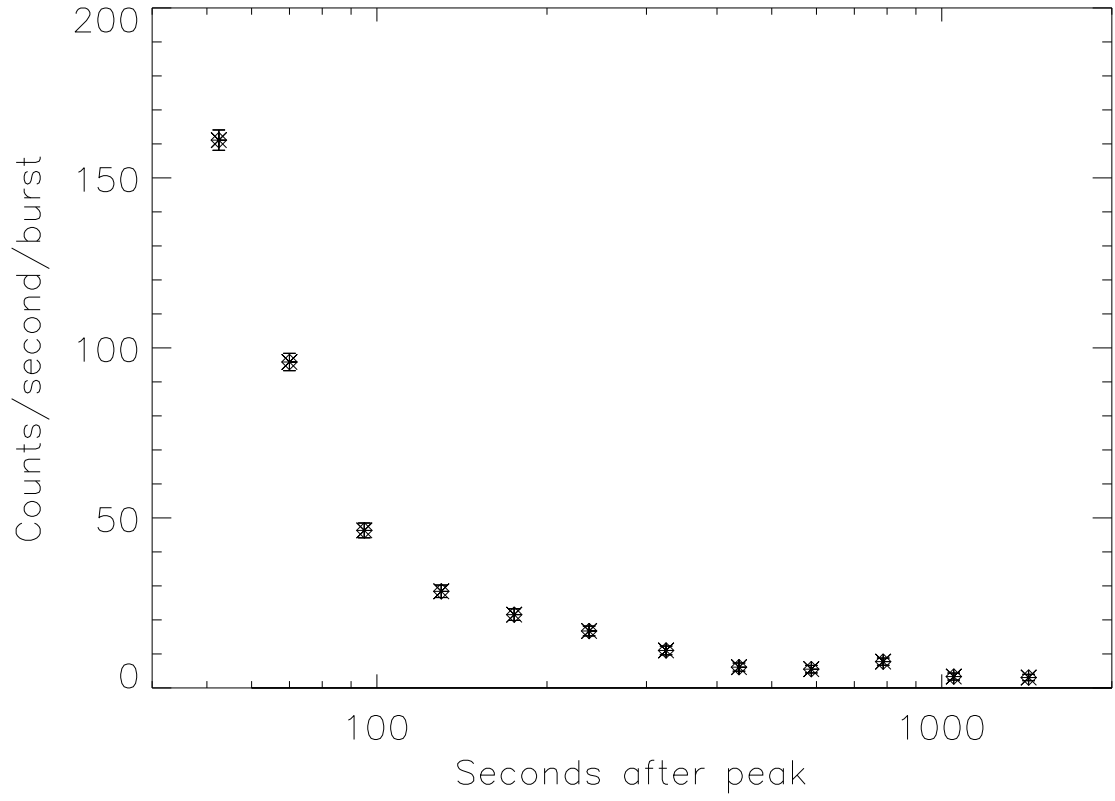}
\caption
{Lightcurve for the 400 long ($>$ 2s) summed and background-subtracted 
BATSE bursts after peak alignment, with peak time suppressed,  summed energies.
\label{fig:lin_ltails}}
\end{figure}

To date, no afterglow emission has been seen from short bursts, but
no such bursts have been detected in prompt X-rays
with the SAX Wide-Field Cameras and consequently no follow-up observations
have been performed for well-localized short events.
It is not known whether the lack of detected afterglows from short bursts
is a function of their being deficient in prompt X-ray emission, whether
instrumental biases result in poor triggering efficiency for
these events using the SAX GRBM/WFC system,   or whether
the smaller numbers of short GRBs have allowed them to elude detection 
thus far.  In this analysis, there does appear to be some excess in the tens of seconds 
after the summed, aligned peaks of the short
GRBs in Figure~\ref{fig:lin_ltails},  consistent in onset time with the 
predictions of ~\citet{apj99:sp} 
that the early afterglow of short GRBs will be separated from
the prompt emission by dozens of seconds, while in longer bursts the
afterglow will overlap the initial burst.  
The flux at a given frequency at the onset of the afterglow 
is determined by an initial Lorentz factor.  It is conceivable that
in these short bursts there is an afterglow signal at gamma-ray energies 
which is just at the sensitivity of the analysis presented in this paper,
but the evidence is more tantalizing than conclusive.

Long-lived tail emission does seem to be a common
attribute of long BATSE GRBs.
It is not clear whether tail components are associated with
the prompt emission or
the afterglow, but the combined light curves from various ensembles
of events suggests they are somewhat independent
of the duration and intensity of the initial burst of gamma rays.  
Owing to the overlap in time between the tail and the main emission
in individual bursts like GRB 980923~\citep{apj99:giblin} 
evidence for 
the distinct natures of the two components comes mainly from spectral differences
between them.  
The most exciting discovery from the analysis of the 19 individual
bursts presented here is the tail following the bright burst
GRB~991216 for which afterglow
candidates were observed at lower wavelengths.
V. Connaughton et al. (in preparation) find  
a preliminary spectral analysis of this burst
confirms the hypothesis of distinct components.
Detection of high-energy afterglow emission
coincident with low energy afterglow observations in individual
bursts expands the energy range over which energy-frequency
temporal dependencies predicted in blast-wave
models for the production of afterglow emission can
be probed~\citep{mnras98:wrm}. 
Although BATSE can no longer provide this high-energy platform,
HETE and Swift will be  operational
during the next few years.  The observations of GRBs 
at optical (Swift), X-ray and gamma-ray energies either
simultaneously or within seconds of a trigger will avoid the time lag
inherent in the pioneering afterglow pursuits of SAX and RXTE
which require time-consuming
spacecraft repointings.  Swift will be sensitive to lower gamma-ray fluxes than
BATSE so that tails of individual bursts will be easier to probe.
While HETE is not as sensitive, its circular orbit facilitates
background subtraction and any long-lived, low-level
emission will be easier to discern. 
Both of these major GRB space missions
will be decisive in answering the questions of the
prevalence and nature of tails in long GRBs and the possible presence
of tails in the short events.    

\begin{acknowledgments}
The author thanks an anonymous referee for constructive comments.
She is grateful to Re'em Sari and to Tim Giblin, Michael Briggs,
Chryssa Kouveliotou, and Matt Scott of the BATSE team
for critical reading of and suggestions for improvements to this paper.
She credits Ralph Wijers with the encouragement and nagging which
spurred the completion of the paper.
The author also acknowledges the National Research Council
postdoctoral fellowship program which supported the research presented here.
\end{acknowledgments}

\clearpage

\clearpage

\clearpage

\clearpage

\clearpage

\clearpage

\clearpage

\clearpage

\begin{table}
\begin{center}
\caption{Tails and afterglows of individual GRBs
\label{tab:ind}}
\centerline{
\begin{tabular}{cccccccccc}\hline
\multicolumn{5}{c}{Tailed} &  \multicolumn{5}{c}{Tailless} \\
\tableline
GRB date & X & O & I & R & GRB date & X & O & I & R \\
\tableline
970111 & N & N & & N & 970508 & Y & Y & Y & Y \\
971024 & & N & & & 971214 & Y & Y & Y & N  \\
980329 & Y & Y & Y & Y & 980611 & N & & & \\
980613 & Y & Y & N & & 990520 &  & N & & \\
980703 & Y & Y & Y & Y & & & & & \\
990506 & Y & N & & &  & &  & & \\
991216 & Y & Y & Y & Y & & & & & \\
\tableline
\end{tabular}}
\end{center}
\end{table}

\end{document}